\documentclass[conference]{IEEEtran}
\IEEEoverridecommandlockouts
\usepackage{cite}
\usepackage{amsmath,amssymb,amsfonts}
\usepackage{graphicx}
\usepackage{textcomp}
\usepackage{xcolor}
\def\BibTeX{{\rm B\kern-.05em{\sc i\kern-.025em b}\kern-.08em
    T\kern-.1667em\lower.7ex\hbox{E}\kern-.125emX}}
    
\usepackage{tikz}
\usetikzlibrary{positioning}
\usepackage{pgfplots}
\usepackage{mathtools}
\usepackage{xparse}
\usepackage{amssymb}
\usepackage{pgfplotstable,booktabs}
\usepackage{graphicx}
\usepackage{subfig}
\usepackage[english]{babel}
\usepackage{romannum}
\usepackage{url}
\usepackage{hyperref}
\usepackage{rotating}
\usepackage{comment}
\usepackage{listings}
\usepackage{float}
\usepackage{stfloats}
\usepackage{filecontents}

\usepackage{listings}
\usepackage{algorithm}
\usepackage{algorithmicx}
\usepackage[noend]{algpseudocode}
\usepackage{multirow}

\DeclarePairedDelimiter\ceil{\lceil}{\rceil}

\begin{document}

\title{Towards High Performance, Portability, and Productivity: Lightweight Augmented Neural Networks for Performance Prediction\\
}

\author{\IEEEauthorblockN{
Ajitesh Srivastava\IEEEauthorrefmark{1}\textsuperscript{\textsection}
Naifeng Zhang\IEEEauthorrefmark{1}\textsuperscript{\textsection}
Rajgopal Kannan\IEEEauthorrefmark{2}
and
Viktor K. Prasanna\IEEEauthorrefmark{1}}

\IEEEauthorblockA{
\IEEEauthorrefmark{1}University of Southern California\\
\{ajiteshs,naifengz,prasanna\}@usc.edu \\
\IEEEauthorrefmark{2}US Army Research Lab-West\\
rajgopal.kannan.civ@mail.mil}

}

\maketitle

\begingroup\renewcommand\thefootnote{\textsection}
\begin{NoHyper}
\footnotetext{Equal contribution}
\end{NoHyper}
\endgroup

\begin{abstract}
Writing high-performance code requires significant expertise in the programming language, compiler optimizations, and hardware knowledge. This often leads to poor productivity and portability and is inconvenient for a non-programmer domain-specialist such as a Physicist. More desirable is a high-level language where the domain-specialist simply specifies the workload in terms of high-level operations (e.g., matrix-multiply(A, B)), and the compiler identifies the best implementation fully utilizing the heterogeneous platform. For creating a compiler that supports productivity, portability, and performance simultaneously, it is crucial to predict the performance of various available implementations (variants) of the dominant operations (kernels) contained in the workload on various hardware to decide (a) which variant should be chosen for each kernel in the workload, and (b) on which hardware resource the variant should run. To enable the performance prediction, we propose lightweight augmented neural networks for arbitrary combinations of kernel-variant-hardware. A key innovation is utilizing the mathematical complexity of the kernels as a feature to achieve higher accuracy. These models are compact to reduce training time and fast inference during compile-time and run-time. Using models with less than 75 parameters, and only 250 training data instances, we are able to obtain a low MAPE of 3\%, significantly outperforming traditional feed-forward neural networks on 48 kernel-variant-hardware combinations. We further demonstrate that our variant-selection approach can be used in Halide implementations to obtain up to 1.7x speedup over Halide's auto-scheduler.
\end{abstract}


\section{Introduction}

    
With various heterogeneous technologies emerging today, there have been unprecedented opportunities for accelerating applications. Application-specific integrated circuits (ASICs) ~\cite{ASICbook} provide highly specialized implementations but require expertise in implementation and are specialized for one application. On the other hand, CPUs, GPUs, and FPGAs provide more flexibility and are easier to program, but are much slower compared to ASICs. Providing the flexibility in applications and ease of implementation while reaching the speedup offered by ASICs has been the focus of many recent works~\cite{Kuon:2009:QEG:1816457,1240929}. Even writing a CPU/GPU code to get most out of available hardware requires programming expertise, hardware knowledge, and time. Further, that optimized code may not be ``portable'', i.e., may not work well on a different platform. Finally, a domain-specialist such as a physicist is expected to know the operations involved in their workload, but not the details of their highly-optimized implementations. This is important for ``productivity'', i.e., implementing the desired workflow with few lines of code, not worrying about the code optimizations.

With the objective of achieving high performance, portability, and productivity,
we are building a compiler that executes a high-level domain-specific language on heterogeneous platforms aligned with recent DARPA projects~\cite{SDH}. The user will write a high-level code that can be broken down in high-level operations (matrix multiplication, convolution, etc.), we call kernels. The user only specifies the operation with the inputs such as \texttt{matrix-multiply(A, B)} without worrying about the optimized implementation of the actual multiplication, thus enabling high \textbf{productivity}. It is the compiler's job to automatically identify how to best execute this code by distributing the kernels among the available hardware configurations on the platform. 

In order to identify a high \textbf{performance} execution plan, the compiler should be able to predict the performance of a kernel on various hardware resources. This enables the following decisions: \textit{(i) Variant-selection:} A compiler may have several variants implementing the kernel on the same hardware in its library with potentially different performances, e.g., Boost library vs Eigen library for matrix multiplication. The variation may also come from setting certain parameters in the implementation that affect the runtimes, such as compilation flags and other tunable parameters of the implementation. Given the input, which variant should be selected? \textit{(ii) Mapping to hardware:} The workload is a collection of possibly interdependent kernels. Each kernel can be mapped to various available hardware resources (CPUs, GPUs, etc.). For each kernel-hardware pair, there may be a different kernel variant that is optimal. Having accurate kernel performance models is crucial for these decisions. We acknowledge that our approach to designing this compiler is not suited for compiling arbitrary low-level code as we rely on already available implementations of certain kernels. However, the kernels chosen in the paper dominate the runtime of many workflows including machine learning. In fact, our chosen kernels cover $>$80\% of the workflows~\cite{SDH_Hongkuan} in the DARPA SDH program~\cite{SDH,PAPPA}. We emphasize that predicting the execution time is more useful than simply knowing the better variant or hardware resource for individual kernels. For instance, suppose, we want to execute two matrix multiplications that do not have any data dependencies on a platform containing a CPU and a GPU. The first one involves matrices of size $100$ and the second of size $10000$. While the first multiplication alone may be faster on GPU, it should still be scheduled on the CPU so that the GPU is available for the second which is the larger multiplication.


To enable \textbf{portability}, the compiler must support learning performance models of execution times $T(K_i, H_j)$ on arbitrary platforms, where $K_{i}$ is an arbitrary kernel implemented on an arbitrary hardware $H_j$. We do not assume any access to hardware profilers or details of the kernel implementation. The kernel implementations on various hardware are treated as black-boxes and we can only manipulate the inputs to the implementations. This makes our approach easily extensible when a new implementation of a kernel is added to the library.
These performance models can be trained during compiler installation by generating benchmark datasets for each kernel (along with its variants) on the available hardware. To make this feasible, the models must be lightweight so that they can learn quickly with small training data without overfitting. 
Once the models are trained, the compiler will be ready for scheduling kernels at compile-time. The prediction may also be needed at runtime. The exact input to the kernel may not be known at compile-time, and therefore, the mapping decisions (which variant to select and where to run) will have to be made dynamically at runtime. Making the models compact is necessary to ensure that they do not constitute a significant portion of the runtime.
Here, we build performance models for four ubiquitous kernels~\cite{SDH}
found in common workflows (i) Matrix-Matrix Multiplication, (ii) Matrix-Vector Multiplication, (iii) Matrix Convolution, and (iv) Max-Pooling. We propose a novel approach called Augmented Neural Network (\textbf{NN+C}) which is extremely lightweight and utilizes the time complexity function to perform execution time prediction.

\noindent\textbf{Key Contributions:} Our key contributions are as follows.
    \begin{itemize}
        \item We propose novel lightweight neural network models for kernel performance prediction on CPUs and GPUs.
        \item We demonstrate that the lightweight models are portable to more than 48 kernel-variant-hardware combinations. Results from 48 combinations have been discussed that include 4 kernels each of which has 2 variants on 3 CPUs each and 2 variants on 2 GPUs each. 
        \item We demonstrate that our models achieve low MAPE of $3\%$ with a small training set in a short amount of training time outperforming traditional feed-forward networks for all 48 kernel-variant-hardware combinations.
        \item We demonstrate that our performance models can be used to identify the best implementation of a kernel where thousands of variants can exist with significantly different runtimes. Specifically, for Halide~\cite{ragan2013halide} implementation of Blur filter our approach results in up to $1.7\times$ speed up over Halide auto-scheduler.
    \end{itemize}
    
\section{Related Work}

Most existing works focus on predicting the performance of the whole specific workload. Huang et. al.~\cite{NIPS2010_4145} use sparse polynomial regression to predict the execution time of arbitrary programs. In~\cite{ipek2005approach}, a neural network is used to predict the execution time of a workload. On the other hand,~\cite{smith1998predicting} proposes feature selection from workloads to identify similar applications for which the runtimes are known and predicting the runtime for the given application using mean or linear regression. These approaches are limited to one or similar applications and will require retraining for every application, and thus is not a scalable approach. Further, it is not clear what type of workloads will result in good predictions and if a similar approach can be ported to other hardware. Instead, we perform predictions at coarse-level building blocks of a program on various hardware. If a compiler can predict performance at coarse level operations (kernels such as matrix multiply) on available hardware, it can make mapping decisions accordingly. For this, we consider four kernels that are dominant in many other workloads. Therefore, instead of being tied to a particular workflow, our approach applies to many, such as the entire class of deep learning workloads. 

Other existing works~\cite{mendis2018ithemal,konstantinidis2017quantitative} rely on the instruction set architecture or hardware-specific metrics, which can potentially be used to predict kernel (instead of workload) performance. However, this would require explicit knowledge of the hardware and corresponding profilers, and thus will reduce portability. Our approach enables a black-box treatment of the kernels and allows prediction without knowing the specific architecture or implementation details. Table~\ref{tab:comparison} summarizes the works closest to ours. 
Although we do not compare our approach against the above-mentioned works quantitatively, as they are for different objectives, we do show comparison against their chosen machine learning models (neural networks and linear regression) and show that our lightweight augmented neural networks achieve superior accuracy.
Finally, our work is different from~\cite{wu2015gpgpu} as they focus on performance prediction of hardware using hardware profiling instead of the performance of dominant operations.
    
    \begin{table}[!htbp]
        \centering
        \caption{Distinction from related works}
        \begin{tabular}{|c|c|c|}
            \hline
            Approach & Workload Coverage & Portability \\
            \hline
            Workload-specific~\cite{NIPS2010_4145,ipek2005approach,smith1998predicting} & Low & N/A\\
            ISA/Hardware specific~\cite{mendis2018ithemal,konstantinidis2017quantitative} & High & Low/Medium \\
            Our work & Medium & High \\
            \hline
        \end{tabular}
        \label{tab:comparison}
    \end{table}
    

\section{Proposed Approach} 

    \textbf{Problem Definition:} For each operation on an arbitrary platform with arbitrary implementations, given corresponding inputs, find a lightweight model that accurately predicts the execution time using a small amount of training time. 
    
    To solve this problem, we propose the Augmented Neural Network (NN+C). 
    The key idea of NN+C is utilizing known mathematical function $f(K,H)$ as an extra input to NN. For example, in Matrix-Matrix Multiplication, besides using basic features such as matrix dimensions, matrix density as inputs, we calculate the number of total operations during Matrix-Matrix Multiplication. Therefore, $f(K,H) = m\times{n}\times{k}$. $f(K,H)$ for Matrix-Vector Multiplication, Matrix Convolution, and Max-Pooling is also calculated similarly. The lightweight aspect enables fast decision making during compile-time as well as run-time. These augmented neural networks provide the flexibility to incorporate any tunable parameter available for the kernel and the hardware.
    
\subsection{Neural Network Structure} 
    The structure of NN+C is shown in Figure \ref{fig:nnstruct}. Inputs to the neural network are 
    \begin{enumerate}
    	\item known mathematical function $f(K,H)$
    	\item kernel parameters $K_{i}$, such as input matrix dimension and matrix density
    	\item hardware/code-optimization parameters $H_{j}$, for example, how many threads are used in the multi-threaded implementation and other controllable features that may affect the runtime such as compilation flags
    \end{enumerate}
    Our augmented neural network contains less than two hidden layers. The output layer has one node, which is the predicted execution time. The number of nodes in hidden layers varies given different kernels and different inputs, resulting in different models for each kernel. Further, models for a given kernel differ for CPU and GPU due to different inputs: in CPU we use multi-threading and take the number of threads as input. Thus, for example, for four kernels mentioned above and two hardware configurations, this results in eight different neural network structures. However, the structure of the models remains the same irrespective of the implementation of the kernel (e.g., different software libraries), and the type of CPU or the type of GPU (e.g., Intel or AMD). In this case, only the weights in the neural network that are learned during training will change. 

\tikzset{%
  every neuron/.style={
    circle,
    draw,
    minimum size=0.5cm
  },
  neuron missing/.style={
    draw=none, 
    scale=2,
    text height=0.333cm,
    execute at begin node=\color{black}$\vdots$
  },
}
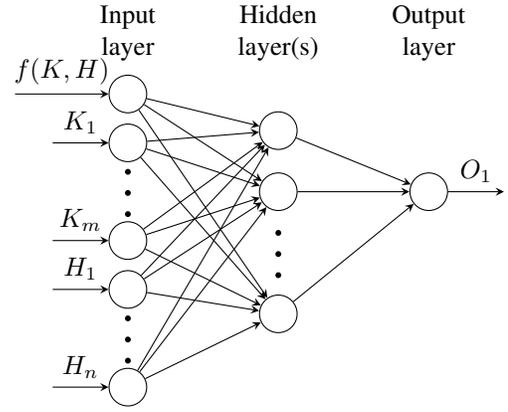
\begin{figure}
    \centering
\begin{tikzpicture}[x=1cm, y=0.65cm, >=stealth] 
   

\foreach \m/\l [count=\y] in {1,2,missing,3,4,missing,5}
  \node [every neuron/.try, neuron \m/.try] (input-\m) at (0,2.5-\y) {};

\foreach \m [count=\y] in {1,2,missing,3}
  \node [every neuron/.try, neuron \m/.try ] (hidden-\m) at (2,2-\y*1.25) {};

\foreach \m [count=\y] in {1}
  \node [every neuron/.try, neuron \m/.try ] (output-\m) at (4,0.5-\y) {};

\draw [<-] (input-1) -- ++(-1.5,0)
    node [above, midway] {$f(K,H)$};
\draw [<-] (input-2) -- ++(-1,0)
    node [above, midway] {$K_1$};    
\draw [<-] (input-3) -- ++(-1,0)
    node [above, midway] {$K_m$};    
\draw [<-] (input-4) -- ++(-1,0)
    node [above, midway] {$H_1$}; 
\draw [<-] (input-5) -- ++(-1,0)
    node [above, midway] {$H_n$};    
    

\foreach \l [count=\i] in {1}
  \draw [->] (output-\i) -- ++(1,0)
    node [above, midway] {$O_\l$};

\foreach \i in {1,...,5}
  \foreach \j in {1,...,3}
    \draw [->] (input-\i) -- (hidden-\j);

\foreach \i in {1,...,3}
  \foreach \j in {1}
    \draw [->] (hidden-\i) -- (output-\j);

\foreach \l [count=\x from 0] in {Input}
  \node [align=center, above] at (\x*2,2) {\l \\ layer};

\foreach \l [count=\x from 1] in {Hidden}
  \node [align=center, above] at (\x*2,2) {\l \\ layer(s)};
  
\foreach \l [count=\x from 2] in {Output}
  \node [align=center, above] at (\x*2,2) {\l \\ layer};

\end{tikzpicture}
\caption{Structure of Augmented Neural Network (NN+C)}
\label{fig:nnstruct}
\end{figure}

    \subsection{Model Inputs}
    
    \paragraph{Matrix-Matrix Multiplication($A_{m,n}\times{B_{n,k}}$)}
    Inputs are the dimensions of the matrices $m$, $n$, and $k$, densities of matrix $A$ ($d_{1} = \frac{\text{number of non-zero entries}}{m\times{n}}$) and of matrix $B$ ($d_{2}$), and the number of threads we utilize during multi-threading on CPU, $N_{thd}$, which is an extra input for operations on CPU and not present for GPU. We augment the neural network with $c = f(K, H)$, i.e.,  roughly the total number of operations in the kernels. In this case, $c=m\times{n}\times{k}$.
    
    \paragraph{Matrix-Vector Multiplication ($A_{m,n}\times{B_{n,1}}$)}
        Inputs are $m$, $n$, $d$, $c$, $N_{thd}$ as defined above. $m$ and $n$ are dimensions of matrix A. $d$ is the density of matrix $A$ and the density of vector $B$ is set as 1. $c$ is the number of operations, $c = m\times{n}$. $N_{thd}$ is the number of threads.
    
    \paragraph{Matrix Convolution($A_{m,n}*{B_{r,r}}$)}
    Inputs are $m$, $n$, $d$, $c$, $N_{thd}$ as defined above, and $r$ is the dimension of square matrix B. $d$ is the density of matrix $A$ and the density of square matrix $B$ is set as 1. The number of operations is given by $c = (m-r+1)\times (n-r+1)\times r^{2}$. $N_{thd}$ is the number of threads.
    
    \paragraph{Max-Pooling ($A_{m,n}*{B_{s,s}}$)}
    Inputs are $m$, $n$, $d$, $c$, $N_{thd}$ as defined above, and $s$ is the dimension of square matrix B. $d$ is the density of matrix $A$ and the density of square matrix $B$ is set as 1. The number of operations is given by $c = \ceil*{\frac{n}{s}}\times{\ceil*{\frac{m}{s}}}\times{s^{2}}$. $N_{thd}$ is the number of threads.
    
    
    
   
\section{Experiments}
\label{sec:exp}
    
    \subsection{Platforms and Optimizations}
    
    To demonstrate portability of our models we conducted our experiments on five platforms: Intel(R) Xeon(R) CPU E5-2650 v2 @ 2.60GHz (\textbf{Xeon}), Intel(R) Core i7-8750H CPU @ 2.20GHz (\textbf{I7}), Intel(R) Core i5-7360U CPU @ 2.30GHz (\textbf{I5}), NVIDIA Tesla K40c (\textbf{Tesla}) and NVIDIA Quadro K420 (\textbf{Quadro}). 
    
    To perform the kernel operations on CPU, we used the Eigen library and Boost library in C++. Eigen/Dense, Eigen/Sparse, uBLAS/matrix, and uBLAS/matrix\_sparse are used to optimize dense and sparse matrix in each kernel. Multi-threading was also used in Eigen to vary the number of threads. However, it is difficult to vary the number of threads without heavily changing the code structure in the Boost library. Owing to our black-box approach, we used a single thread in the Boost library. Among our platforms, Xeon has 16 cores, 32 threads; I7 has 12 cores, 24 threads; and I5 has 2 cores, 4 threads. For all operations on GPU, we used two implementations of CUDA to optimize, one through global memory and one through shared memory. This results in 10 implementations of each kernel: 2 variants of 3 CPUs and 2 variants on 2 GPUs. We published our code for reproduciblility\footnote{\url{https://github.com/Naifeng/Augmented-Neural-Network}}. 
    
    \subsection{Datasets}
    
    We measured the performance of four kernels on each platform: Matrix-Matrix Multiplication (\textbf{MM}), Matrix-Vector Multiplication (\textbf{MV}), Matrix Convolution (\textbf{MC}) and Max-Pooling (\textbf{MP}). Other kernels such as LU decomposition and Blur filter were also evaluated, but their results have been omitted due to brevity. For each kernel-variant-hardware combination (there are 40 such combinations), we generated 500 instances of data, where 250 instances were used to train the model and 250 instances to test. Each data instance was generated randomly with ranges of parameters as described in Table~\ref{tab:paramgen}. While the experiments may be conducted with a different set of ranges, we chose these ranges as they are common sizes for deep learning workflows. Since we use multi-threading on CPU, all operations on CPU take an extra input $N_{thd}$, which is randomly generated between 1 to the maximum threads supported by the given platforms. 

\begin{table}[ht]
\caption{Parameters for data generation}
\begin{center}
\begin{tabular}{|c|c|}
    \hline
    \multirow{3}{*}{Matrix-Matrix Multiplication} & $m,n,k\in \{1,2,3, \dots, 1024\}$\\
    &$d_{1} \in \{{{{1,\frac{1}{2},\frac{1}{4},\dots, \frac{1}{2^{\log_{2} m\times{n}}}}}}\}$\\
    &$d_{2} \in \{{{{1,\frac{1}{2},\frac{1}{4},\dots, \frac{1}{2^{\log_{2} n\times{k}}}}}}\}$ \\
    \hline
    \multirow{2}{*}{Matrix-Vector Multiplication} & $m,n\in \{1,2,3, \dots, 1024\}$\\
            &$d\in \{{{{\frac{1}{2},\frac{1}{4},\frac{1}{8},\dots, \frac{1}{2^{\log_{2} m\times{n}}}}}}\}$\\
    \hline
    \multirow{3}{*}{Matrix Convolution} & $r \in \{3,5,7\}$\\
            & $m,n\in \{r,r+1,r+2, \dots, 1024\}$\\
            & $d\in \{{{{1,\frac{1}{2},\frac{1}{4},\dots, \frac{1}{2^{\log_{2} m\times{n}}}}}}\}$\\
    \hline
    \multirow{4}{*}{Max-Pooling} & $r \in \{2,3,4,5\}$\\
            & $s \in \{1,2\}$\\
            & $m,n\in \{r,r+1,r+2, \dots, 1024\}$\\
            & $d\in \{{{{1,\frac{1}{2},\frac{1}{4},\dots, \frac{1}{2^{\log_{2} m\times{n}}}}}}\}$\\
    \hline
\end{tabular}
\end{center}
\label{tab:paramgen}
\end{table}
    
    \subsection{Models}
    \label{sec:models}
    Our augmented neural networks are built under the TensorFlow framework. Each model is kept under 75 parameters to maintain lightweight and a short training time. All models have 2 dense layers and use ReLU as the activation function. We use Adam as the optimizer \cite{kingma2014adam}, with learning rate varying between 0.01, 0.0001, and 0.0001. The loss function is chosen to be mean squared error. Each epoch included training with a full batch.
    The number of parameters of each model as well as its average training time is shown in Table \ref{table:1}.

    \begin{table}[H]
        \begin{center}
        \caption{Number of parameters and average training times}
        \label{table:1}
        \begin{tabular}{ |c|c|c|c|c| } 
        \hline
        & MM      & MV      & MC     & MP \\  
        \hline
    CPU & 64, 19s & 50, 18s & 73, 6s & 73, 6s \\
    GPU & 41, 19s & 73, 6s  & 50, 8s & 73, 7s \\
        \hline
        \end{tabular}
        \end{center}
    \end{table}
    
    \subsection{Baselines}
    \label{sec:baselines}
    We compare our method against four baselines: (1) Neural Network (\textbf{NN}). NN is the same as the implementation of NN+C except that NN does not take the number of operations as an extra input. (2) Constant (\textbf{C}). In C, we only take the number of operations as input and try to predict execution time using linear regression. (3) Augmented Linear Regression (\textbf{LR+C}). We take the same inputs as NN+C but use linear regression in LR+C. (4) Augmented Non-Linear Regression (\textbf{NLR+C}). In NLR+C we take the same inputs as NN+C but use the random forest regression \cite{breiman2001random}. Random forest based regression has been demonstrated to be competitive in performance prediction \cite{dahinden2011improved, hutter2014algorithm}.

    \subsection{Evaluation Metrics}
    
    We used mean absolute percentage error (MAPE) to evaluate the predictions $\{\hat{t}_1, \hat{t}_1, \dots, \hat{t}_N\}$ obtained by the baselines and our models w.r.t. the ground truth $\{t_1, t_2, \dots, t_N\}$:
    \begin{equation}
    \label{eq:mape}
        MAPE = \frac{100}{N}\sum_i \frac{|t_i - \hat{t_i}|}{t_i}.
    \end{equation}
    
    By the definition of MAPE, a small misprediction ($|t_i - \hat{t_i}|$) might lead to a exceptionally high MAPE (up to 5000\%) if the true runtime $t_{i}$ is minute. Those extreme MAPE values skew the average despite most of the predictions being accurate. Thus, we introduced a threshold at the 30\% of the testing data, ranking from the lowest runtime to the highest runtime. Overall, the average runtime of testing data below the threshold is 13\% of the average runtime of all the testing data, but these low runtime data instances contribute approximately 80\% of the overall MAPE. For example, when analyzing NN+C performance on MC on GPU, MAPE of the lowest 30\% is 128\% and the MAPE of the highest 70\% is 15\%, whereas the overall MAPE is 49\%. Therefore, in reporting MAPE, we drop 30\% of testing data with the lowest runtime for a more precise assessment of models' performance.  

\section{Demonstration of Variant-Selection}
\label{sec:var}

As a crucial application of our performance prediction approach, we demonstrate that it can be used to pick the best variant for a given kernel, i.e., picking the best available code among several options. Possible scenarios include choosing between a CPU and a GPU implementation and identifying compilation flags that will be best suited for the kernel. To show the variant selection capability of our approach, we choose a scenario where the number of variants can be extremely high. Further, we choose two kernels different than the four discussed thus far to show the generalizability of our approach.

We consider the Blur kernel (\textbf{Blur}) and Fast Fourier transform (\textbf{FFT}) kernel implemented in Halide~\cite{ragan2013halide}. A Halide code decouples the functional program from its execution ``schedule'' that determines various aspects of the execution such as the ordering of the loops, degree of unrolling loops, and vectorization strategy. The schedule description can be considered as a combination of shape (feature space) and parameters. For instance, 

\begin{center}
\texttt{blur\_y.tile(x, y, xi, yi, 128, 256)}
\end{center}

defines two dimensions of the shape and the parameters 128 and 256 determine the tunable parameters along these dimensions. Changing the schedule does not affect the output of the code, but it may significantly affect the runtime. Therefore, each schedule generates a variant of the same kernel, and our task is to identify the best variant to use.

\subsection{Model Inputs}

We train our compact augmented neural networks with inputs representing the schedule features. This allows us to quickly estimate runtimes of the code with various schedule parameters without actually executing the code.  Halide provides an auto-scheduler~\cite{mullapudi2016} that attempts to identify the best schedule itself. We run the auto-scheduler to identify the shape/feature space and ignore the suggested parameters. Within this feature space we generate candidate schedules $S = \{s_1, s_2, s_3, \dots, s_N\}$, where each $s_i$ is a vector of parameters, and find $s = \arg \min_i P_{}(s_i)$, where $P_{}(s_i)$ represents the predicted runtime given by schedule $s_{i}$. For kernels that Halide auto-scheduler are not applicable to, we identify the feature space based on the provided manually written implementation. 

Input data dimensions $n$ and augmented constant are also fed into the neural network. We augmented $n^{2}$ for Blur and $n\log_{2}n$ for FFT to corresponding variant-selection models given the complexity of Halide implementation of both kernels \cite{li1989memory}. 

\subsection{Platforms and Optimizations}

We conducted variant-selection experiments on five platforms: Xeon, I7, I5, Tesla, and Quadro. We used Halide to implement the kernel operations. More experiment settings of variant-selection can be found at our published code\footnote{\url{https://github.com/Naifeng/Variant-Selection}}.

\subsection{Datasets}

We evaluated two kernels: Blur and FFT. The performance of Blur is measured on five platforms. Given that there is no existing GPU schedule of FFT provided by Halide, we only conducted experiments of FFT on three CPU platforms. To demonstrate that our models are able to identify the best implementation among numerous existing variants, we generated thousands of data instances and restricted the training set to consist of 250 instances to maintain portability. 

The following is a piece of code from the implementation of Halide Blur on CPU. Each of $s_{1}$, $s_{2}$, $s_{3}$ and $s_{4}$ resides in a \texttt{.split()} function and serves as a split factor. The inner loop runs from zero to the split factor and the outer loop runs from zero to the extent required by the first argument divided by the split factor \cite{ragan2013halide}. Thus, a combination of $\{s_{1}$, $s_{2}$, $s_{3}$, $s_{4}\}$ defines a candidate schedule and different schedules have significantly different runtimes. We varied each parameter extensively to generate a candidate set. The schedule given by Halide auto-scheduler is $\{$8, 256, 128, 8$\}$. 

\begin{lstlisting}
{
    Var x = blur_x.args()[0];
    blur_x
        .compute_at(blur_y, x_o)
        .split(x, x_vo, x_vi, s1)
        .vectorize(x_vi);
}
{
    Var x = blur_y.args()[0];
    Var y = blur_y.args()[1];
    blur_y
        .compute_root()
        .split(x, x_o, x_i, s2)
        .split(y, y_o, y_i, s3)
        .reorder(x_i, y_i, x_o, y_o)
        .split(x_i, x_i_vo, x_i_vi, s4)
        .vectorize(x_i_vi)
        .parallel(y_o)
        .parallel(x_o);
}
\end{lstlisting}

According to Halide implementation rules and current supports (e.g., Halide only supports limited input dimensions for FFT), we varied parameters as described in Table \ref{tab:paramgen_vs}. For Blur and FFT on CPU, we generated 1000 data instances for each input data dimension, resulting in 6000 instances and 4000 instances, respectively. For Blur on GPU, we exhaustively generated all possible combinations, that is, 1176 instances. 

\begin{table}[ht]
\caption{Parameters for data generation}
\begin{center}
\begin{tabular}{|c|c|}
    \hline
    \multirow{4}{*}{Blur (CPU)} 
    & $n\in \{2^{10},2^{11},2^{12},\dots, 2^{15}\}$ \\ 
    & $s_{1},s_{2}\in \{2,4,8,\dots,1024\}$ \\
    & $s_{3}\in \{2,4,8,\dots,s_{2}\}$ \\
    & $s_{4}\in \{2,4,8,\dots,s_{3}\}$ \\
    \hline
    \multirow{3}{*}{Blur (GPU)} 
    & $n\in \{2^{10},2^{11},2^{12},\dots, 2^{15}\}$ \\ 
    & $s_{1}\in \{2,4,8,16\}$ \\
    & $s_{2}, s_{3} \in \{1,2,4, \dots,64\}$ \\
    \hline
    \multirow{3}{*}{FFT (CPU)}
    & $n\in \{2^{4},2^{5},2^{6},2^{7}\}$ \\ 
    & $s_{1}\in \{2,4,8,\dots,2^{n-1}\}$ \\
    & $s_{2}, s_{3}, s_{4}, s_{5}, s_{6}\in \{2,4,8,\dots,2^{n}\}$ \\
    \hline
\end{tabular}
\end{center}
\label{tab:paramgen_vs}
\end{table}

\subsection{Models}

Augmented neural networks used for variant-selection is the same as models described in Section \ref{sec:models} except that all models used for variant-selection have 3 dense layers. The number of parameters of each model as well as its average training time is shown in Table \ref{table:params}.

\begin{table}[H]
        \begin{center}
        \caption{Number of parameters and average training times}
        \label{table:params}
        \begin{tabular}{ |c|c|c|c|c| } 
        \hline
        & Blur     & FFT       \\  
        \hline
    CPU & 71, 18s & 67, 12s  \\
    GPU & 66, 7s & N/A  \\
        \hline
        \end{tabular}
        \end{center}
    \end{table}

\subsection{Baselines}

We compare our method against four baselines identical to baselines described in Section \ref{sec:baselines}: (1) NN, (2) C, (3) LR+C, and (4) NLR+C. In addition, for Blur on CPU, we compare our variant-selection approach with the Halide auto-scheduler to show the overall improvement. For Blur on GPU, due to the fact that Halide does not have a stable auto-scheduler to generate a GPU schedule, we compare our variant-selection results with the average runtime among the runtime of all candidate schedules. Similarly, since current Halide auto-scheduler is not capable of scheduling a complicated pipeline such as FFT, we compare our results with the average runtime among the runtime of all candidate schedules for FFT on CPU. 

\begin{figure*}[!ht]
\centering
    \includegraphics[width=0.3\textwidth,trim=4 30 4 4,clip]{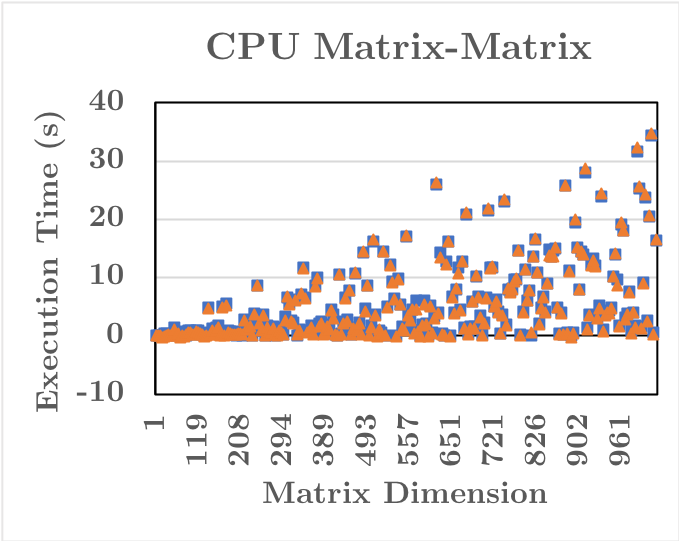}
    \includegraphics[width=0.3\textwidth,trim=4 30 4 4,clip]{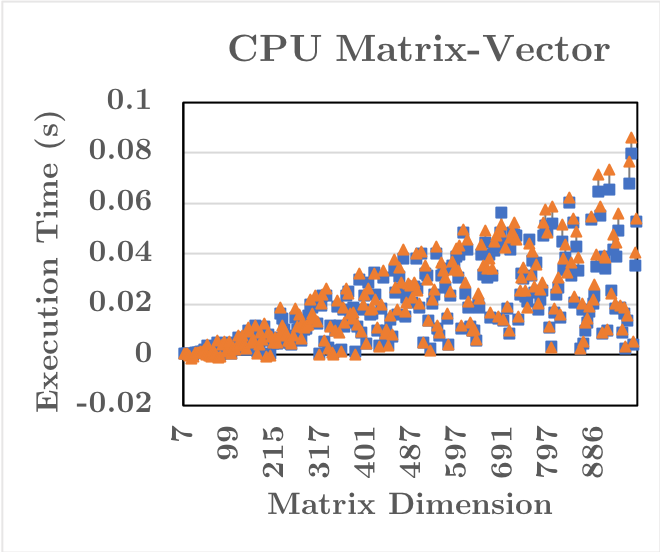}
    \includegraphics[width=0.3\textwidth,trim=4 30 4 4,clip]{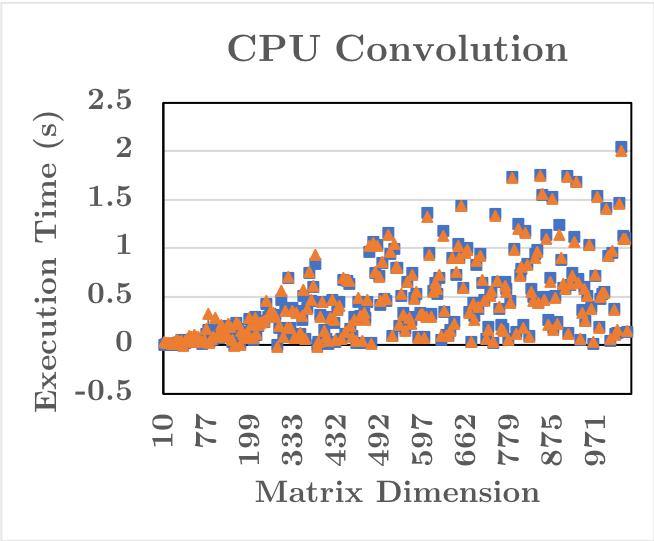} \\
    \includegraphics[width=0.3\textwidth,trim=4 30 4 4,clip]{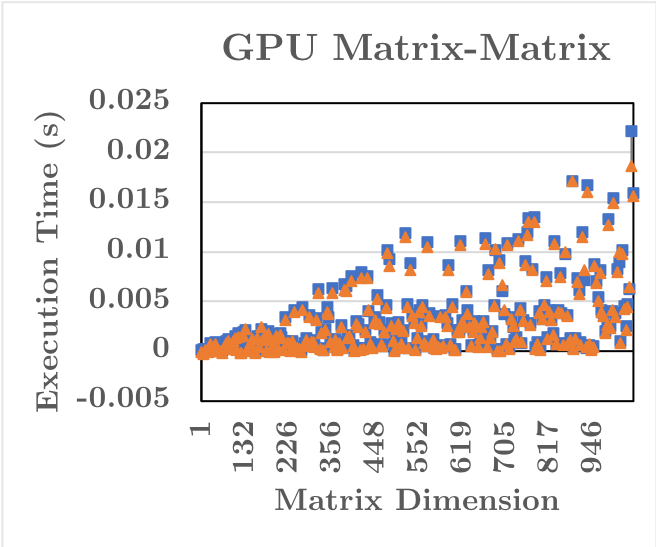}
    \includegraphics[width=0.3\textwidth,trim=4 30 4 4,clip]{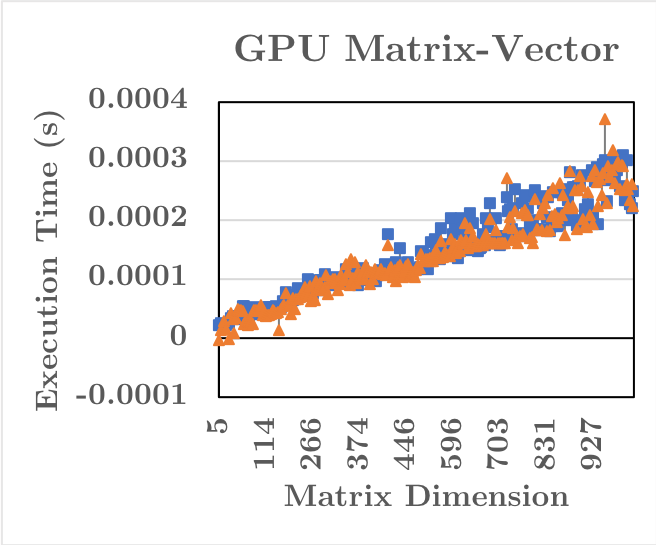}
    \includegraphics[width=0.3\textwidth,trim=4 30 4 4,clip]{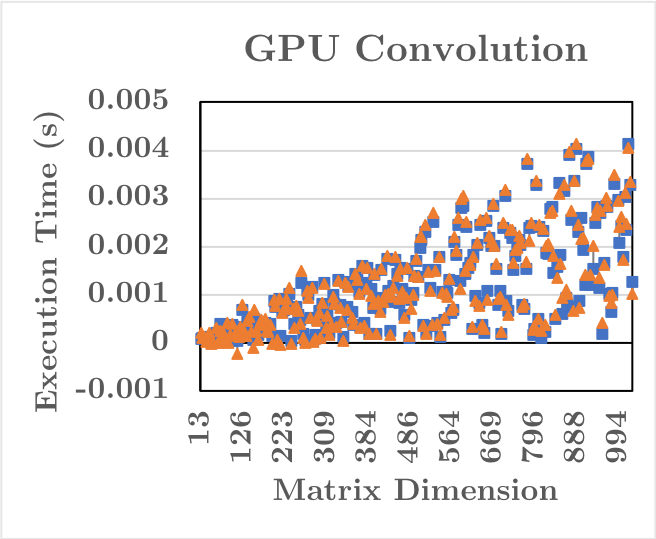} \\
    \includegraphics[width=0.3\textwidth,trim=4 30 4 4,clip]{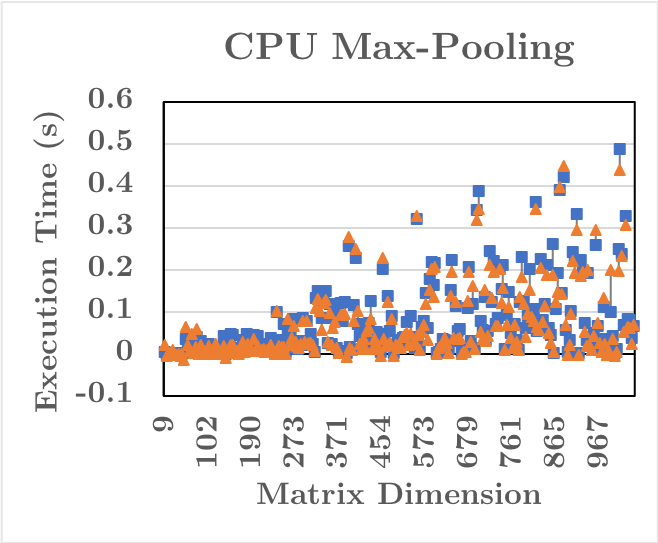} 
    \includegraphics[width=0.3\textwidth,trim=4 30 4 4,clip]{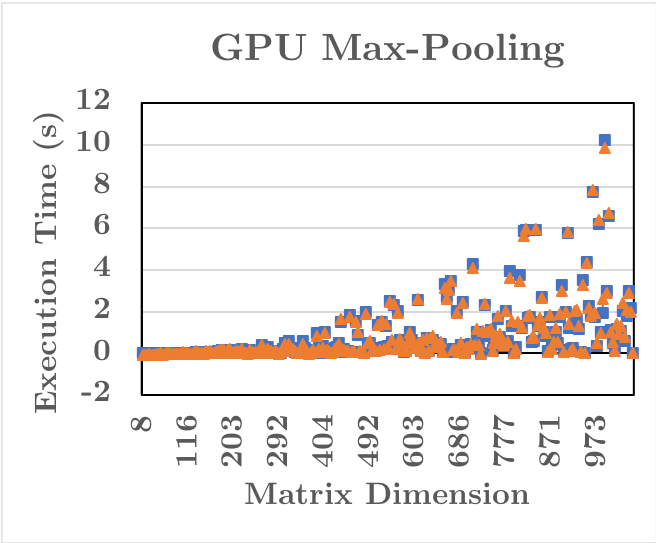}
    \\
        \includegraphics[scale=0.3]{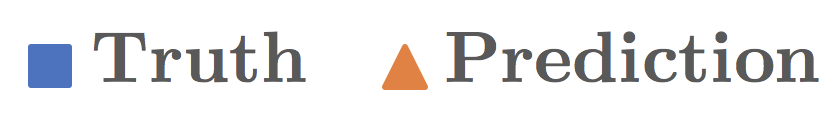}
    \caption{Performance predictions of four kernels using NN+C}

    \label{fig:predCPU}
\end{figure*}

\subsection{Evaluation Metrics}
We used MAPE and Spearman's rank correlation coefficient ($\rho$) to evaluate the predictions $\{\hat{t}_1, \hat{t}_1, \dots, \hat{t}_N\}$ obtained by the baselines and our models w.r.t. the ground truth $\{t_1, t_2, \dots, t_N\}$. MAPE is defined in Equation (\ref{eq:mape}). $\rho$ is defined as 
\begin{equation}
\rho = 1-\frac{6 \sum_{i=1}^{N} d_{i}^{2}}{N\left(N^{2}-1\right)}
\end{equation}
where $d_{i}=|\operatorname{rank}(t_{i})-\operatorname{rank}(\hat{t}_i)|$. $\operatorname{rank}(t_{i})$ is the rank of $t_{i}$ among $\{t_1, t_2, \dots, t_N\}$, ranking from the lowest value to the highest value. $\operatorname{rank}(\hat{t}_{i})$ is the rank of $\hat{t}_{i}$ among $\{\hat{t}_1, \hat{t}_1, \dots, \hat{t}_N\}$, ranking from the lowest value to the highest value. $\rho$ ranges from $-1$ to $1$. $\rho$ of $1$ indicates a perfect positive correlation of two variables' ranks and $\rho$ of $-1$ indicates a perfect negative correlation of two variables' ranks. The closer $\rho$ is to zero, the weaker the correlation between the ranks.


\begin{table*}[!ht]
  \begin{center}
    \caption{Prediction MAPE of Matrix-Matrix Multiplication}
    \label{tab:MM}
    \begin{tabular}{|c|c|c|c|c|c|c|c|c|c|c|}
      
      \hline
      & \multicolumn{6}{c|}{CPU}  & \multicolumn{4}{c|}{GPU} \\
      \cline{2-11}
      & \multicolumn{3}{c|}{Eigen}  & \multicolumn{3}{c|}{Boost} & \multicolumn{2}{c|}{CUDA\textsubscript{Global Memory}} & \multicolumn{2}{c|}{CUDA\textsubscript{Shared Memory}}\\
      \cline{2-11}
      & Xeon & I7 & I5 & Xeon & I7 & I5 & Tesla & Quadro & Tesla & Quadro \\
      \hline
      NN+C
      & \textbf{14\%} & \textbf{23\%} & \textbf{8\%} & \textbf{7\%} & \textbf{27\%} & \textbf{6\%} & \textbf{7\%} & \textbf{5\%} & \textbf{8\%} & \textbf{8\%}\\
      \hline
      NN 
      & 29\% & 31\% & 26\% & 20\% & 35\% & 19\% & 23\% & 13\% & 18\% & 16\% \\
      \hline
      C
      & 39\% & 34\% & 28\% & 8\% & 34\% & 7\% & 9\% & 9\% & 10\% & 10\% \\
      \hline
      LR+C
      & 44\% & 31\% & 33\% & 8\% & 34\% & 7\% & 8\% & 8\% & \textbf{8\%} & \textbf{8\%} \\
      \hline
      NLR+C
      & 23\% & 24\% & \textbf{8\%} & 9\% & 33\% & 7\% & 10\% & 10\% & 18\% & 19\% \\

      \hline
    \end{tabular}
  \end{center}
\end{table*}

\begin{table*}[!ht]
  \begin{center}
    \caption{Prediction MAPE of Matrix-Vector Multiplication}
    \label{tab:MV}
    \begin{tabular}{|c|c|c|c|c|c|c|c|c|c|c|}
      \hline
    & \multicolumn{6}{c|}{CPU}  & \multicolumn{4}{c|}{GPU} \\
      \cline{2-11}
      & \multicolumn{3}{c|}{Eigen}  & \multicolumn{3}{c|}{Boost} & \multicolumn{2}{c|}{CUDA\textsubscript{Global Memory}} & \multicolumn{2}{c|}{CUDA\textsubscript{Shared Memory}}\\
      \cline{2-11}
      & Xeon & I7 & I5 & Xeon & I7 & I5 & Tesla & Quadro & Tesla & Quadro \\
      
      \hline
      NN+C
      & \textbf{21\%} & \textbf{21\%} & \textbf{25\%} & \textbf{11\%} & \textbf{8\%} & \textbf{9\%} & \textbf{7\%} & \textbf{7\%} & \textbf{7\%} & \textbf{6\%} \\
      \hline
      NN
      & 22\% & 24\% & 29\% & 14\% & 11\% & 12\% & \textbf{7\%} & 8\% & \textbf{7\%} & 9\% \\
      \hline
      C
      & \textbf{21\%} & 22\% & \textbf{25\%} & 12\% & \textbf{8\%} & \textbf{9\%} & 23\% & 23\% & 11\% & 10\% \\
      \hline
      LR+C
      & \textbf{21\%} & 22\% & 26\% & 12\% & \textbf{8\%} & \textbf{9\%} & \textbf{7\%} & \textbf{7\%} & \textbf{7\%} & \textbf{6\%} \\
      \hline
      NLR+C
      & 26\% & 25\% & 27\% & 12\% & \textbf{8\%} & \textbf{9\%} & 29\% & 28\% & 21\% & 22\% \\
      
      \hline
    \end{tabular}
  \end{center}
\end{table*}


\begin{table*}[!ht]
  \begin{center}
    \caption{Prediction MAPE of Matrix Convolution}
    \label{tab:MC}
    \begin{tabular}{|c|c|c|c|c|c|c|c|c|c|c|}
      \hline
    & \multicolumn{6}{c|}{CPU}  & \multicolumn{4}{c|}{GPU} \\
      \cline{2-11}
      & \multicolumn{3}{c|}{Eigen}  & \multicolumn{3}{c|}{Boost} & \multicolumn{2}{c|}{CUDA\textsubscript{Global Memory}} & \multicolumn{2}{c|}{CUDA\textsubscript{Shared Memory}}\\
      \cline{2-11}
      & Xeon & I7 & I5 & Xeon & I7 & I5 & Tesla & Quadro & Tesla & Quadro \\
      
      \hline
      NN+C 
      & \textbf{8\%} & \textbf{21\%} & \textbf{4\%} & \textbf{30\%} & \textbf{20\%} & \textbf{13\%} & \textbf{10\%} & \textbf{15\%} & \textbf{17\%} & \textbf{19\%} \\

      \hline
      NN
      & 9\% & 22\% & 7\% & 50\% & 30\% & 30\% & 16\% & \textbf{15\%} & 22\% & \textbf{19\%} \\

      \hline
      C
      & 27\% & 40\% & 22\% & 48\% & 44\% & 40\% & 30\% & 30\% & 42\% & 42\% \\

      \hline
      LR+C
      & 15\% & 32\% & 13\% & 46\% & 38\% & 37\% & 15\% & \textbf{15\%} & 29\% & 30\%\\
      
      \hline
      NLR+C
      & 18\% & 32\% & 7\% & \textbf{30\%} & 32\% & 24\% & 17\% & 17\% & 21\% & 21\%\\
      
      \hline
      
    \end{tabular}
  \end{center}
\end{table*}


\begin{table*}[!ht]
  \begin{center}
    \caption{Prediction MAPE of Max-Pooling}
    \label{tab:MP}
    \begin{tabular}{|c|c|c|c|c|c|c|c|c|c|c|}
      \hline
    & \multicolumn{6}{c|}{CPU}  & \multicolumn{4}{c|}{GPU} \\
      \cline{2-11}
      & \multicolumn{3}{c|}{Eigen}  & \multicolumn{3}{c|}{Boost} & \multicolumn{2}{c|}{CUDA\textsubscript{Global Memory}} & \multicolumn{2}{c|}{CUDA\textsubscript{Shared Memory}}\\
      \cline{2-11}
      & Xeon & I7 & I5 & Xeon & I7 & I5 & Tesla & Quadro & Tesla & Quadro \\
      
      \hline
      NN+C 
      & \textbf{23\%} & \textbf{13\%} & \textbf{14\%} & \textbf{27\%} & \textbf{12\%} & \textbf{14\%} & \textbf{14\%} & \textbf{8\%} & \textbf{25\%} & \textbf{27\%} \\
      \hline
      NN
      & 32\% & 20\% & 22\% & 36\% & 20\% & 34\% & 32\% & 32\% & 40\% & 47\% \\
      \hline
      C
      & 67\% & 37\% & 43\% & 81\% & 41\% & 47\% & 93\% & 95\% & 40\% & 28\% \\
      \hline
      LR+C
      & 50\% & 26\% & 27\% & 63\% & 31\% & 33\% & 75\% & 77\% & 40\% & 28\% \\
      \hline
      NLR+C
      & 25\% & \textbf{13\%} & 17\% & \textbf{27\%} & 13\% & 18\% & \textbf{14\%} & 14\% & 31\% & 29\% \\
      \hline
      
    \end{tabular}
  \end{center}
\end{table*}

\section{Results}
\label{sec:results}
\subsection{Kernel Performance Prediction}
\label{sec:unconstrained} 

   Figures \ref{fig:predCPU} shows a visualization of using NN+C to predict kernel performance on two platforms. We choose the results of I5 and Tesla to represent the results on CPU and GPU, respectively. We pick matrix dimension as x-axis, plotting against execution time in seconds to visualize prediction. A line joining two points in the plot indicates the corresponding prediction and ground truth. Note that very few points have a significant misprediction. Tables~\ref{tab:MM}, \ref{tab:MV}, \ref{tab:MC}, and \ref{tab:MP} quantify these results using MAPE.

    For all five kernels using any implementation, NN+C produces the lowest MAPE in predictions. 
    Ranking from the highest accuracy (lowest MAPE) on average to the lowest is (1) NN+C, (2) NLR+C, (3) NN, (4) LR+C, and (5) C.  
    On average, NN+C outperforms traditional NN by a margin of 8\% and outperforms the second-best approach NLR+C by 5\%. LR+C has a good prediction for kernels on GPU. Performance of C on all platforms is worst among all kernels expect MV. Overall, NN+C predicts more accurately for kernels on GPU than those on CPU, achieving on average a low MAPE of 12\% and 16\%, respectively. 
    

    We report the average error in MAPE for the four kernels and the two hardware classes (CPU, GPU) in Table~\ref{tab:MAPE}. For each kernel, MAPE was aggregated over all hardware and variants. For each hardware class, MAPE was aggregate over all the kernels, variants, and specific devices. We show the comparison of NN+C against traditional NN. NN+C significantly outperforms NN in almost all cases. In fact, for MM, MAPE for NN+C is less than half of that of NN suggesting that the traditional neural network is far inferior than our augmented neural network.
    
    \begin{table}[H]
    \centering
            \caption{Aggregated MAPE of NN+C vs. NN}
    \begin{tabular}{ |c|c|c|c|c|c|c| } 
    \hline
    & MM & MV & MC & MP & CPU & GPU\\
    \hline
    NN+C & 11\% & 12\% & 16\% & 18\% & 16\% & 12\%\\ 
    \hline
    NN & 23\% & 14\% & 22\% & 32\% & 24\% & 20\%\\ 
    \hline
    \end{tabular}
    \label{tab:MAPE}
\end{table}

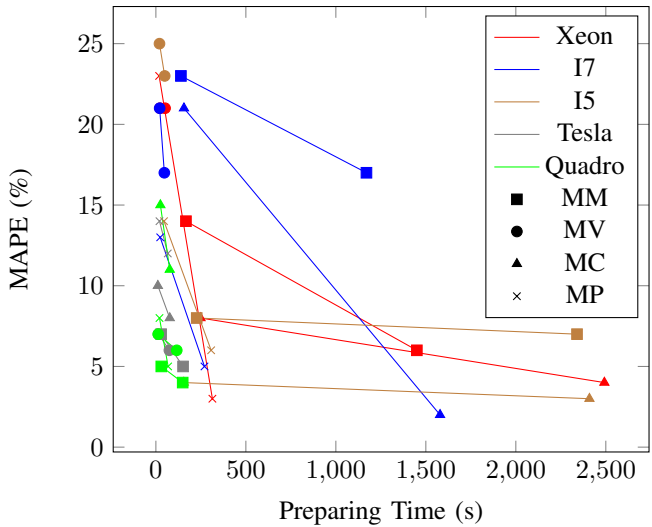
\begin{figure}[t]
\centering
\begin{tikzpicture}
\begin{axis}[
    title={},
    xlabel={Preparing Time (s)},
    ylabel={MAPE (\%)},
    legend entries={Xeon,I7,I5,Tesla,Quadro,MM,MV,MC,MP},
    width=0.48\textwidth
]

\addlegendimage{no markers,red}
\addlegendimage{no markers,blue}
\addlegendimage{no markers,brown}
\addlegendimage{no markers,gray}
\addlegendimage{no markers,green}
\addlegendimage{only marks,mark=square*,black}
\addlegendimage{only marks,mark=*,black}
\addlegendimage{only marks,mark=triangle*,black}
\addlegendimage{only marks,mark=x,black}

\addplot+[color=red, mark options={mark=square*, fill=red},solid] table [x=MM_T, y=MM_E, col sep=comma] {sl_Xeon.csv};
\addplot+[color=red, mark options={mark=*, fill=red},solid] table [x=MV_T, y=MV_E, col sep=comma] {sl_Xeon.csv};
\addplot+[color=red, mark options={mark=triangle*, fill=red},solid] table [x=MC_T, y=MC_E, col sep=comma] {sl_Xeon.csv};
\addplot+[color=red, mark options={mark=x, fill=red},solid] table [x=MP_T, y=MP_E, col sep=comma] {sl_Xeon.csv};

\addplot+[color=blue, mark options={mark=square*, fill=blue},solid] table [x=MM_T, y=MM_E, col sep=comma] {sl_I7.csv};
\addplot+[color=blue, mark options={mark=*, fill=blue},solid] table [x=MV_T, y=MV_E, col sep=comma] {sl_I7.csv};
\addplot+[color=blue, mark options={mark=triangle*,fill=blue},solid] table [x=MC_T, y=MC_E, col sep=comma] {sl_I7.csv};
\addplot+[color=blue, mark options={mark=x,fill=blue},solid] table [x=MP_T, y=MP_E, col sep=comma] {sl_I7.csv};

\addplot+[color=brown, mark options={mark=square*,fill=brown},solid] table [x=MM_T, y=MM_E, col sep=comma] {sl_I5.csv};
\addplot+[color=brown, mark options={mark=*,fill=brown},solid] table [x=MV_T, y=MV_E, col sep=comma] {sl_I5.csv};
\addplot+[color=brown, mark options={mark=triangle*,fill=brown},solid] table [x=MC_T, y=MC_E, col sep=comma] {sl_I5.csv};
\addplot+[color=brown, mark options={mark=x,fill=brown},solid] table [x=MP_T, y=MP_E, col sep=comma] {sl_I5.csv};

\addplot+[color=gray, mark options={mark=square*,fill=gray},solid] table [x=MM_T, y=MM_E, col sep=comma] {sl_Tesla.csv};
\addplot+[color=gray, mark options={mark=*,fill=gray},solid] table [x=MV_T, y=MV_E, col sep=comma] {sl_Tesla.csv};
\addplot+[color=gray, mark options={mark=triangle*,fill=gray},solid] table [x=MC_T, y=MC_E, col sep=comma] {sl_Tesla.csv};
\addplot+[color=gray, mark options={mark=x,fill=gray},solid] table [x=MP_T, y=MP_E, col sep=comma] {sl_Tesla.csv};

\addplot+[color=green, mark options={mark=square*,fill=green},solid] table [x=MM_T, y=MM_E, col sep=comma] {sl_Quadro.csv};
\addplot+[color=green, mark options={mark=*,fill=green},solid] table [x=MV_T, y=MV_E, col sep=comma] {sl_Quadro.csv};
\addplot+[color=green, mark options={mark=triangle*,fill=green},solid] table [x=MC_T, y=MC_E, col sep=comma] {sl_Quadro.csv};
\addplot+[color=green, mark options={mark=x,fill=green},solid] table [x=MP_T, y=MP_E, col sep=comma] {sl_Quadro.csv};

\end{axis}
\end{tikzpicture}
\caption{Performance comparison between Lightweight Models and Unconstrained Models} 
\label{fig:comp}
\end{figure}

    \paragraph{Unconstrained Augmented Neural Networks} To enable fast inference, our models are kept extremely lightweight -- less than 75 weights. Also, we only generate 500 data instances for each kernel-variant-hardware combination, out of which 250 are used to train our models. In order to assess how much of the performance is compromised due to these restrictions, we build similar NN+C models with more parameters and generate a larger dataset with 5000 data instances (2500 instances are used to train and 2500 to test). Figure \ref{fig:comp} illustrates the comparison between lightweight models and unconstrained models in terms of error. Overall, MAPE achieved by lightweight NN+C is 14\% and by unconstrained NN+C is 9\%. Specifically, on CPU, using unconstrained NN+C, MM, MV, MC, and MP have a decrease on average MAPE of 5\%, 2\%, 8\%, and 12\%, respectively. On GPU, using unconstrained NN+C, MM, MV, MC, and MP have a decrease on average MAPE of 1.5\%, 1\%, 3\%, and 2.5\%, respectively. However, accuracy comes at the cost of increased model size and the overall time as summarized in Table~\ref{table:2}. 
    \begin{table}[H]
    \centering
    \caption{Preparing time increase and model size increase}
         \label{table:2}
\begin{tabular}{|c|c|c|c|c|}
\hline
    & MM            & MV           & MC            & MP \\ \hline
CPU & 9.31x, 2.13x & 2.30x, 2.12x & 11.59x, 2.21x & 10.24x, 2.48x                      \\ \hline
GPU & 5.08x, 8.80x  & 7.07x, 2.34x & 4.28x, 2.12x  & 3.35x, 2.62x 
\\ \hline
\end{tabular}
\end{table}
    
    The preparing time (training data generation time plus model training time) on average of lightweight NN+C on CPU is 104s and that of unconstrained NN+C is 1040s, which is 10x of lightweight NN+C. The preparing time on average of lightweight NN+C on GPU is 20s and that of unconstrained NN+C is 97s, 4.85x of lightweight NN+C. In addition to the preparing time loss, model size is significantly reduced. Model size reduction is most evident in MM on GPU, compared to unconstrained NN+C, lightweight NN+C model is 8.80x smaller. The rest of the models are downsized by 2.29x on average. 
    
    If the training and inference time is not constrained, then one can use our unconstrained (larger and more accurate) augmented models. However, we envision that lightweight models may be necessary due to the following reasons:
    (a) at compile-time many kernels need to be evaluated: consider VGG16 inference that requires $>$1M 2D-Convolutions. At a given layer, there can be $>$100K 2D-Convolutions, each of which may have different execution times not only due to heterogeneous hardware but also due to different sparsity. This number of convolutions will multiply with the factor of the number of parallel image classification pipelines. (b) Some decisions may have to be made at runtime: some kernels may be only instantiated at runtime, which is the only time performance prediction inference has to be performed. In such scenarios, the inference time should be as minimal as possible to avoid an significant impact on the total runtime.
    
\begin{figure}[!ht]
    \centering
    \begin{tabular}{c}
    \includegraphics[width=0.4\textwidth]{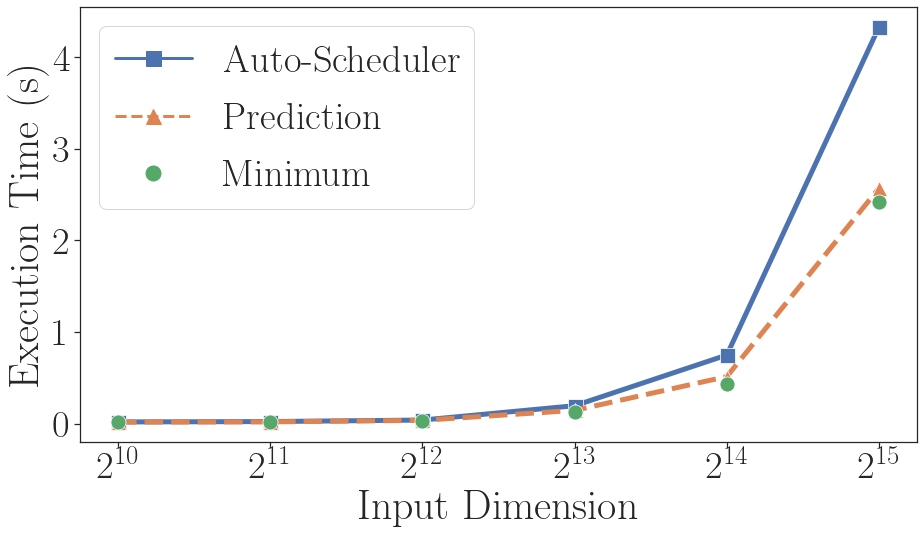} \\
    (a) Halide Blur (CPU)\\
    \\
    \end{tabular}
    \begin{tabular}{c}
    \includegraphics[width=0.4\textwidth]{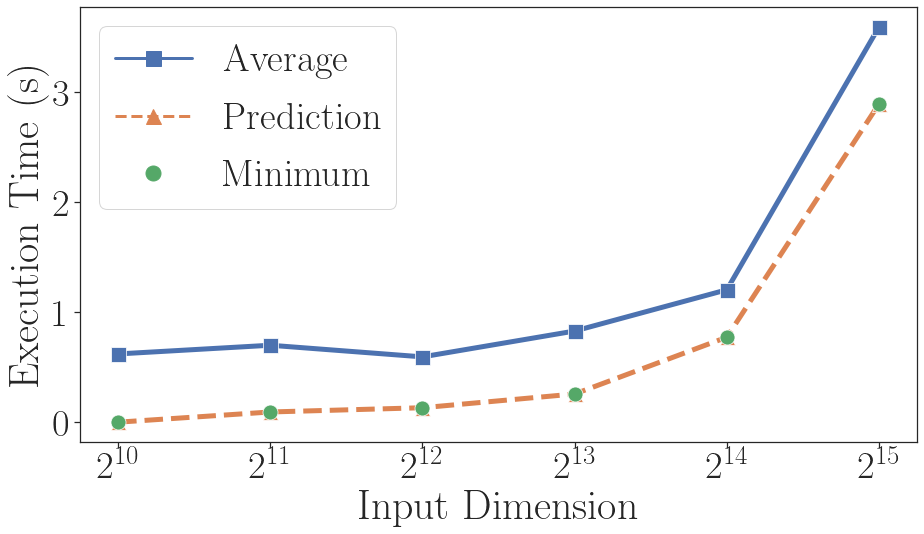} \\ 
    (b) Halide Blur (GPU)\\
    \\
    \end{tabular}
    \begin{tabular}{c}
    \includegraphics[width=0.4\textwidth]{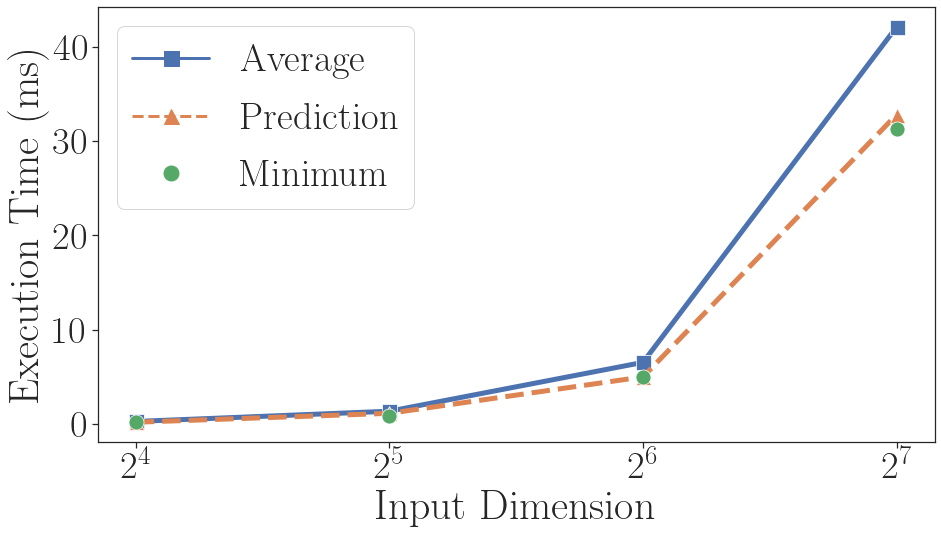} \\
    (c) Halide FFT (CPU) \\
    \\
    \end{tabular}
    \caption{Runtime comparison of variants obtained from baseline and our approach}
    \label{fig:halide_speedup}
\end{figure}

\subsection{Variant-Selection}

Figure~\ref{fig:halide_speedup} shows the comparison of execution times for varying input sizes of two kernels. Our predicted best schedule produces a runtime close to the true best schedule within the candidate set in all cases. Further, Figure~\ref{fig:halide_speedup}(a) shows that using our predictions, we were able to outperform Halide's auto-scheduler, getting up to $1.7\times$ speedup in kernel Blur on CPU. As for Blur on GPU, we obtained up to $223.5\times$ speedup compared to a randomly selected schedule on a small input size ($2^{10}$), see Figure~\ref{fig:halide_speedup}(b), and among all input sizes, we were able to obtain a speedup of at least $1.24\times$. Shown in Figure~\ref{fig:halide_speedup}(c), we obtained up to $1.5\times$ speedup compared to a randomly selected schedule of Halide FFT on CPU.

\begin{figure}[H]
    \centering
    \includegraphics[width = 0.48\textwidth]{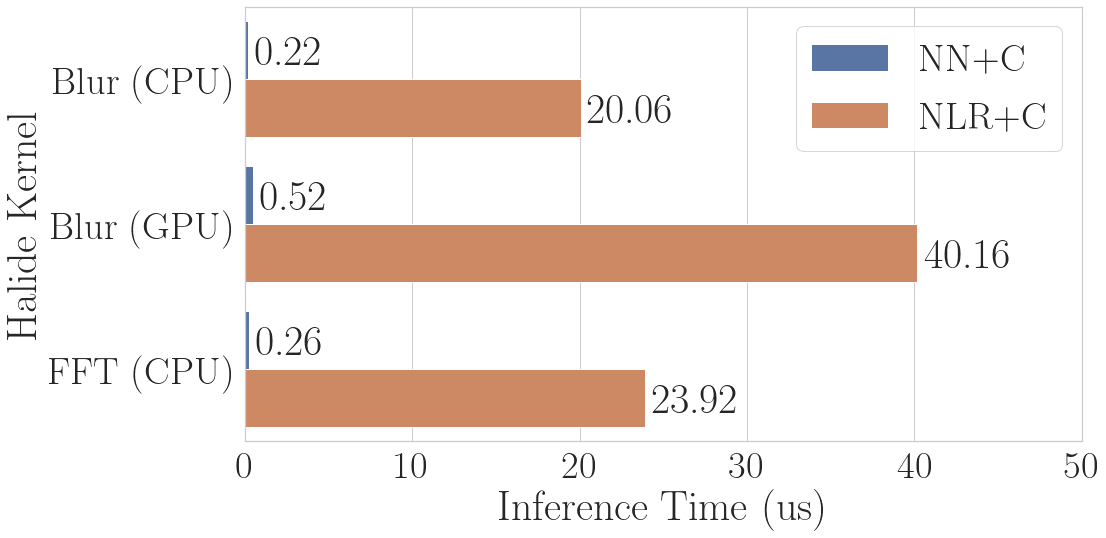}
    \caption{Inference time comparison of NN+C and NLR+C}
    \label{fig:blur_NLR}
\end{figure}

Note that MAPE varies among different train-test splits and training process. The MAPE value shown in Table \ref{tab:mape_rho} is the best (lowest) MAPE obtained by our methods and the baselines on the Halide kernels. The table also shows the Spearman's rank correlation coefficient. Since the main objective is to select the best variant which requires the ability to correctly rank the variants, this is the primary metric of comparison. We observe that our approach NN+C obtains the highest rank correlation in the majority of the cases. NLR+C has a higher rank correlation for Halide Blur while having much worse MAPE. On the other hand, for Halide FFT, NLR+C obtains identical rank coefficients with NN+C and better MAPEs. However, NLR+C has a much higher inference time, and therefore likely to hinder the execution of the actual application if used by the runtime component of a compiler, as we argued in the end of Section~\ref{sec:unconstrained}. Figure~\ref{fig:blur_NLR} shows the comparison of inference times between our lightweight NN+C and NLR+C. It is noteworthy that the inference time of NLR+C is more than $75\times$ of that of our approach.

\begin{table*}[bt]
    \begin{center}
    \caption{Prediction MAPE and Spearman's coefficient}
    \label{tab:mape_rho}
        \begin{tabular}{|c|c|c|c|c|c|c|c|c|c|c|}
        \hline
        & \multicolumn{5}{c|}{Halide Blur} & \multicolumn{3}{c|}{Halide FFT} \\
        \cline{2-9}
        & \multicolumn{3}{c|}{CPU}  & \multicolumn{2}{c|}{GPU} & \multicolumn{3}{c|}{CPU} \\
        \cline{2-9}
        & Xeon & I7 & I5 & Tesla & Quadro & Xeon & I7 & I5 \\
        \hline
        NN+C
        & \textbf{50\%}, 0.91  & \textbf{23\%}, \textbf{0.95} & \textbf{28\%}, \textbf{0.97}   & \textbf{8\%}, \textbf{0.99}     & \textbf{22\%}, \textbf{0.97}    & 8\%, \textbf{0.99}     & 14\%, \textbf{0.97}    & 3\%, \textbf{1} \\
        \hline
        NN
        & 72\%, 0.87  & 25\%, 0.94 & 40\%, 0.91   & 12\%, 0.98    & 29\%, 0.94    & 11\%, 0.98    & 19\%, 0.95  & 17\%, 0.95 \\
        \hline
        C
        & 1140\%, 0.76& 59\%, 0.82 & 167\%, 0.93  & 84\%, 0.73    & 64\%, 0.85    & 66\%, 0.86    & 33\%, 0.97  & 30\%, 0.97 \\
        \hline
        LR+C
        & 1687\%, 0.66& 93\%, 0.82 & 411\%, 0.84  & 106\%, 0.89   & 62\%, 0.87    & 44\%, 0.97    & 32\%, 0.92   & 26\%, 0.90 \\
        \hline
        NLR+C
        & 150\%, \textbf{0.93} & 39\%, 0.94 & 43\%, \textbf{0.97}   & 10\%, \textbf{0.99}    & 23\%, \textbf{0.97}    & \textbf{3\%}, \textbf{0.99}     & \textbf{11\%}, \textbf{0.97}   & \textbf{2\%}, \textbf{1} \\
        \hline
        \end{tabular}
    \end{center}
\end{table*}

\section{Conclusion}

We have proposed a novel lightweight augmented neural network (NN+C), to predict kernel performance on CPUs and GPUs. Our approach is designed in support of creating compilers with high productivity, portability, and performance. To show that our models are portable to different platforms with different implementations, we have evaluated our model on several CPUs and GPUs with multiple optimizations, resulting in a total of 48 kernel-variant-hardware combinations. Our models significantly outperformed the baselines including standard neural network. We have shown that our approach can be used to identify the best variants even when the number of variants is extremely high. We do so by demonstrating a $1.7\times$ speedup over Halide auto-scheduler.  In future work, we will build prediction models for other popular kernels. These models will be used to perform optimized mapping of kernels in workflows for various heterogeneous platforms.

\section*{Acknowledgement}

This work is supported by the Defense Advanced Research Projects Agency (DARPA) under BAA number HR0011-20-9-0019 and by the National Science Foundation Award number 1911229. Any opinions, findings, and conclusions or recommendations expressed in this material are those of the authors and do not necessarily reflect the views of the sponsors.

\bibliography{ref} 
\bibliographystyle{ieeetr}

\end{document}